\documentclass[twocolumn,showpacs,pra,nofootinbib]{revtex4-1}
\usepackage[T1]{fontenc}
\usepackage[russian,english]{babel}
\usepackage{graphicx}
\usepackage{amsmath}
\usepackage{amssymb}
\usepackage{mathtools}
\usepackage{physics}
\usepackage{color,soul}
\usepackage[colorlinks=true,allcolors=blue]{hyperref}
\usepackage{cancel}

\usepackage{upgreek}

\graphicspath{{./Figs/}}

\begin{document}

\title{Beamstrahlung-enhanced disruption in beam-beam interaction}

\date{\today}
\author{A.~S.~Samsonov}
\email[Corresponding author: ]{asams@ipfran.ru}
\author{E.~N.~Nerush}
\author{I.~Yu.~Kostyukov}
\affiliation{Institute of Applied Physics of the Russian Academy of
	Sciences, 46 Ulyanov St., Nizhny Novgorod 603950, Russia}
\author{M.~Filipovic} 
\author{C.~Baumann}
\author{A.~Pukhov}
\affiliation{Institut f\"ur Theoretische Physik I, Heinrich-Heine-Universit\"at D\"usseldorf, Universit\"atsstr. 1, 40225 D\"usseldorf, Germany}

\begin{abstract}
    The radiation reaction (beamstrahlung) effect on particle dynamics during interaction of oppositely charged beams is studied. It is shown that the beam focusing can be strongly enhanced due to beamstrahlung. An approximate analytical solution of the motion equation including the radiation reaction force is derived. The disruption parameter is calculated for classical and quantum regime of beamstrahlung. The analytical model is verified by QED-PIC simulations. The model for head-on collision of long beams undergoing a number of betatron oscillation during interaction is also developed. It is demonstrated that the beamstrahlung-enhanced disruption effect can play a significant role in future lepton colliders with high-current particle beams.
\end{abstract}

\maketitle

\section{Introduction}
\label{sec.Intro}

A beam-beam interaction phenomenon is a fundamental problem of plasma sciences and high-energy physics. It plays a key role in many astrophysical processes as well as in scientific instruments. Particularly,  particle colliders, which are the main research tool of high-energy physics, are based on head-on collisions of high-energy beams. There are several projects aiming at constructing high-energy lepton colliders with record parameters such as ILC~\cite{ILC}, CLIC~\cite{CLIC}, FACET-II~\cite{FACET} etc. Intense electromagnetic fields are generated at the interaction point thereby making possible manifestation of some strong-field phenomena such as disruption~\cite{hollebeek1981disruption,yokoya1992beam,chen1988disruption}, beamstrahlung~\cite{noble1987beamstrahlung,blankenbecler1987quantum,bell1995quantum}, electron-positron pair production~\cite{chen1989coherent,esberg2014strong} or even effects of nonperturbative strong field quantum electrodynamics (QED)~\cite{yakimenko2019prospect,tamburini2020efficient}.

In ultrarelativistic regime the dynamics of a beam particle is governed mainly by the field of the counter-propagating beam while the field of the own beam can be neglected~\cite{davidson2001physics,katsouleas1990plasma}. In this approximation the Lorentz force acting on the beam particle can be written as
\begin{equation}
\mathbf{F} = q \mathbf{E} + (q/c) \left[ \mathbf{v} \times \mathbf{B}  \right] \simeq \pm m \omega_b^2 \mathbf{r},
\label{Lorentz_force}
\end{equation} 
where $\mathbf{v}$ is the particle velocity, $c$ is the speed of light, $\mathbf{E}$ and $\mathbf{B}$ are the electric and the magnetic fields of the counter-propagating beam, respectively, $q=\pm e$ is the particle charge, $r$ is the distance from the particle to the beam axis, $\omega_b = \left( 4 \pi e^2 n_b/m \right)^{1/2} $ is the electron (positron) plasma frequency, $n_b$ is the density of the counter-propagating beam,  $m$ and $e>0$ are positron's mass and charge, respectively.
The positive sign in Eq.~\eqref{Lorentz_force} refers to the case of $e^- e^-$ collisions, where the net force causes a defocusing of the two colliding beams~\footnote{Of course, the same also holds for $e^+ e^+$ collisions}. On the other hand, the beam particle undergoes transverse betatron oscillations with frequency $\omega_b/ \sqrt{\gamma}$ in the case of $e^- e^+$ collisions~\cite{chen1988introduction,chen1988disruption}. Here, $\gamma$ is the Lorentz factor of the beam particle. The disruption (or pinching) time can be introduced as the time, it takes the particle to reach the beam axis, and it can be estimated (up to numerical factor) as follows
\begin{equation}
T_D = \frac{\sqrt{2\gamma} }{\omega_b} .
\label{td}
\end{equation}  
If the beam length $\sigma_x$ fulfills the condition $\sigma_x/c >  T_D$, then the beam radii are significantly changed during the interaction. The beam distortion in the interaction region can be quantified by the so-called disruption parameter which reads
\begin{equation}
D = D_0 \equiv \frac{\sigma^2_x}{c^2 T_D^2} = \frac{\omega^2_b \sigma^2_x}{2 \gamma c^2 }
\label{d1}
\end{equation} 
for a uniform charge distribution of the beam~\cite{hollebeek1981disruption}. Note that it is $\pi^{-1/2} 2^{3/2}\approx 1.6$ times greater than the disruption parameter for a beam with a Gaussian charge distribution~\cite{hollebeek1981disruption,chen1988introduction}.
The expression for $D$ can be generalized to other beam charge distributions and can be also used to characterize  $e^- e^-$ beam interactions. The beam distortion is negligible if $D \ll 1$ which is favorable for collider operation.

The bending of the particle trajectory at the interaction point is accompanied by synchrotron radiation, which is known under the term beamstrahlung in the collider physics community \cite{blankenbecler1987quantum,chen1988introduction}. The total power of the photon emission  losses depends on the electron (positron) dynamical QED parameter $\chi$ \cite{ritus1985quantum,berestetskii1982quantum,BaierKatkov} 
\begin{eqnarray}
    \label{W1}
    P_\text{rad} &=& \frac{\alpha m^{2}c^{4}}{3 \sqrt{3}\pi\hbar}
    \int_{0}^{\infty}du\frac{4u^{3}+5u^2+4u}{(1+u)^{4}}K_{2/3}\left(\frac{2u}
    {3\chi}\right), \\
    \label{chi}
    \chi &=& \frac{\gamma}{E_S}
    \sqrt{\left(\mathbf{E}+\mathbf{v}\times\mathbf{\mathbf{B}}\right)^{2}-\left(\mathbf{v}\cdot\mathbf{\mathbf{E}}\right)^{2}},
\end{eqnarray}
where $\alpha=e^2/(\hbar c)$ is the fine structure constant, $\hbar$ is the Planck constant, $E_S= m^2 c^3 /(e\hbar)$ is the critical Sauter-Schwinger field~\cite{berestetskii1982quantum}, $K_{\nu}(x)$ and $\Gamma(x)$ are the McDonald function and Gamma function, respectively  \cite{abramowitz1964handbook}. In both the classical ($\chi \ll 1$) and the strong QED ($\chi \gg 1$) limits Eq.~\eqref{W1} can be reduced to simple power-law expressions
\begin{eqnarray}
    \label{W2}
    P_\text{rad}(\chi\ll1) \equiv P_C & = & \frac{2}{3}\, \frac{\alpha m^{2}c^{4}} {\hbar}\chi^2,\\
    \label{W3}
    P_\text{rad}(\chi\gg1) \equiv P_Q & = & 0.37\, \frac{\alpha m^{2}c^{4}}{\hbar} \chi^{2/3}.
\end{eqnarray}
If the beam length is so small, that only few photons are emitted by a single beam particle during the interaction, then the quantum nature of the synchrotron radiation should be taken into account even in the limit $\chi \ll 1$.

In addition to beamstrahlung, other quantum effects are possible at the interaction point such as electron-positron pair production via the decay of beamstrahlung photons in strong electromagnetic fields, trident process {\it etc.} \cite{chen1989coherent,hartin2018strong}. In general, the interplay between radiation of hard-photons and pair production can lead to a very fast growth of the total number of particles~---~an effect known as QED cascade, which has recently drawn a lot of attention~\cite{nerush2007radiation,Bell2008,Nerush2011,Ridgers2012,narozhny2015quantum,Kostyukov2016,grismayer2016laser,grismayer2017seeded,jirka2017qed,luo2018qed,Yuan2018,del2018ion,Lu2018,Luo2018,efimenko2019laser,samsonov2019laser, samsonov2021hydrodynamical}. Such QED cascades can also develop in colliding beam scenarios. It is thereby likely that the cascade development is similar to that observed in laser-solid interactions at scales much less than the laser wavelength~\cite{samsonov2019laser, samsonov2021hydrodynamical} due to similar field configurations in the cascading regions. All in all, beamstrahlung and QED cascades may cause beam distortion  due to energy depletion and in general play a negative role on clean collider operation. In the context of particle physics, colliders are therefore usually designed to mitigate beamstrahlung as much as possible. Understanding the collective effects at the interaction point is nevertheless crucial not only for optimal collider operation but also for high energy density physics. Here, the regime of beam-beam interactions with strong beamstrahlung can be exploited, for example, to produce bright gamma-ray sources or to explore strong-field QED experimentally~\cite{del2019bright,song2021generation,tamburini2020efficient}.

Up to now, analytical beam-beam interaction models considered both disruption and beamstrahlung independently. Here, we advance these works by focusing our study on synergic disruption-beamstrahlung effects and we find modified expressions for the disruption parameter including radiation reaction in both the weak- ($\chi\ll1$) and the strong-field limit ($\chi\gg1$). Strong beamstrahlung causes a fast loss of the particle energy and since the disruption time is proportional to $\sqrt{\gamma}$ [see Eq.~\eqref{td}], the energy loss leads to a reduction of the disruption time. Simultaneously, this results in a growth of the disruption parameter $D$. Vice versa the beam focusing reduces the energy loss as the electromagnetic field strength decreases when approaching the beam axis.

The paper is organized as follows. In Sec.~II we formulate the basic equations describing the physics of the beam-beam interaction. 
In Sec.~III we adopt several assumptions to approximately solve the equations of motion and find an analytical estimate for the beamstrahlung-enhanced disruption parameter.
The interaction of long beams is discussed in Sec.~IV, which may be relevant for the interaction of electron and proton beams. The results of QED-PIC simulations are presented in Sec.~V. They are compared with the model predictions. Section~VI contains a discussion and conclusions. 

Throughout the manuscript, the equations will be given in normalized units. The normalization is mainly determined by the initial plasma frequency of the beam $\omega_b$. Then, time is measured in $ 1/\omega_b$, lengths in $c/\omega_b$, momenta in $mc$, electromagnetic fields in $mc\omega_b/e$, and power in $mc^2\omega_b$.

\section{Problem formulation}
\subsection{General approximations}
 In general, the equations of motion including the radiation reaction force for the ultrarelativistic electron are
\begin{eqnarray}
\frac{d \mathbf{p}}{d\tau} & = & - \mathbf{E} - \frac{\mathbf{p}}{\gamma} \times \mathbf{B}  - P(\chi) \,\frac{\mathbf{p}}{\gamma}, \label{eq.p1} \\
\frac{d \mathbf{r}}{d\tau} & = & \frac{\mathbf{p}}{\gamma},
\end{eqnarray}
where $P$ refers to the power in normalized units.
These equations describe the classical motion of the electron in an electromagnetic field with the radiation reaction effect, where the QED corrections to the radiation reaction force are taken into account in semiclassical approximation \cite{kirk2009pair,bulanov2013electromagnetic,esirkepov2014attractors}. The corrections reduce the impact of radiation reaction when increasing $\chi$.

In order to analytically explore the beam disruption effect during head-on collisions of electron and positron beams, we make additional assumptions. First, as mentioned in Sec.~\ref{sec.Intro}, the self-force generated by an ultrarelativistic beam can be neglected in Eq.~\eqref{eq.p1} since it scales with $\gamma^{-2}$~\cite{davidson2001physics,katsouleas1990plasma}.
Second, it is sufficient to concentrate the study on the transverse dynamics of particles located at the beam's front since they start to feel the collective force from the counter-propagating beam earlier than other particles.
And third, we  further restrict our analysis on particles at the periphery of the beam, i.e. the particles which experience the largest force and thus are more likely to emit photons. As beamstrahlung leads to a decrease of the energy and thus inertia of the particles, it is exactly those particles at the the periphery and at the front of the beams that are expected to experience the strongest disruption. 
The analysis of the motion of such particles is greatly simplified due to the fact that their dynamics is only affected by the unperturbed part of the opposite beam. 
Finally, we assume that the electron and positron beams have identical initial parameters, in which case the beams evolve symmetrically along the propagation axis. In addition, the beams are considered to have cylindrical symmetry. In that case, one can write the beam density distribution as $n(\xi_\pm,r) = n_0 \eta_x(\xi_\pm) \eta_r(r)$, where $n_0$ is the maximum beam density, $\xi_\pm = x\pm \tau$ describes the longitudinal coordinate for beams that travel at the speed of light, and the functions $0 \leq \eta_{r,x} \leq 1$ determine the shape of the density distribution~\footnote{However, the generalization to arbitrary transverse profiles is straightforward.}. The electric field generated by such a beam is mostly transverse. Using Gauss's law, it can be expressed as
\begin{eqnarray}
    E_r & = & \frac{\eta_x(\xi_\pm)}{r}\int\limits_0^r \eta_r(r') r' dr'  = \frac{r_b \eta_x(\xi_\pm)}{2}\, \mathcal{E}(\rho) ,  \label{eq.efield} \\
 \mathcal{E}(\rho) & \equiv & \frac{2}{\rho}\int\limits_0^{\rho} \eta_r(r_b \rho') \rho' d\rho',
\end{eqnarray}
where $\rho = r / r_b$ is the transverse coordinate measured relative to the distance, $r_b$, from the beam axis at which the electric field reaches its maximum. For electrons with $v_x= \text{const} = c $ interacting with the counter-propagating beam, $\xi_+$ is decisive and one finds $\xi_+ = 2 \tau$.

With all that in mind and defining $\eta(\tau) \equiv \eta_x(2\tau)$, the governing equations of motion reduce to
\begin{eqnarray}
    \label{eq.dydt}
    \frac{d^2\rho}{d\tau^2} & = & -\frac{\mathcal{E}(\rho) }{\gamma} \,\eta(\tau),   \\
    \label{eq.dgdt}
    \frac{d\gamma}{d\tau} & = & -P(\chi),  \\
    \label{eq.c2}
    \chi & = & \gamma\, \frac{\mathcal{E}(\rho)}{a_S} \,r_b \eta(\tau).
\end{eqnarray}
Here, $a_S = e E_S / (m c \omega_b) = mc^2 /(\hbar \omega_b)$ represents the Sauter-Schwinger field in normalized units. In the derivation of these equations we assumed that electric and magnetic components of the Lorentz force acting on the particle are almost equal to each other [hence, the factor $1/2$ in Eq.~\eqref{eq.efield} is cancelled], which is valid if $v_x \simeq c \gg v_r$ and $\gamma\gg 1$. This also allows us to assume that the radiation friction force acts mostly along the $x$-axis. Thus, it is not explicitly present in the equation for the transverse coordinate $\rho$.

As mentioned above we will be interested in particles experiencing the largest fields, i.e. particles for which the initial displacement $r_0$ from the beam axis equals $r_b$ and thus $\rho_0 \equiv \rho(\tau=0) = 1$.

\subsection{Timescales}

Before solving Eqs.~\eqref{eq.dydt}~--~\eqref{eq.dgdt}, it is useful to estimate characteristic normalized timescales present in the problem, i.e. the timescale of the electron trajectory $\tau_{D_0}$ and the timescale of the energy loss due to beamstrahlung $\tau_{BS}$
\begin{eqnarray}
    \tau_{D_0} = \sqrt{2\gamma_0}, \\
    \tau_{BS} = \frac{\gamma_0}{P(\chi_0)},
\end{eqnarray}
where $\chi_0 = r_b \gamma_0 \mathcal{E}\left( \rho_0 \right) / a_S $ and $\gamma_0 = \gamma(\tau = 0)$ are the initial values of the $\chi$ parameter and Lorentz-factor of the particles, respectively. Let us also introduce a parameter $\varkappa$ in the following way,
\begin{equation}
    \label{eq.condition}
    \varkappa = \frac{\tau_{D_0}}{\tau_{BS}} = \sqrt{\frac{2}{\gamma_0}}P(\chi_0).
\end{equation}
This parameter determines the regime of the beam-beam interaction. In case $\varkappa \gg 1$, which is explored in Sec.~III, significant energy losses due to beamstrahlung occur on  a timescale much shorter than the time it take the particle to reach the beam axis. In the opposite limit $\varkappa \ll 1$, which is considered in Sec.~IV, it takes many betatron periods for beamstrahlung to significantly decrease the beam energy.

Utilizing the relation between $\gamma_0$ and $\chi_0$, and noting that $P(\chi)\equiv \alpha a_S \Phi(\chi)$, the parameter $\varkappa $ can be also expressed as follows
\begin{equation}
    \label{eq.beta0}
    \varkappa = \alpha \sqrt{ 2 r_b a_S }\; \frac{ \Phi (\chi_0) }{\sqrt{\chi_0}},
\end{equation}
where $\lambdabar_C=\hbar/m_e c$ is the Compton wavelength.
It means that the beamstrahlung effect is determined by two initial parameters of the interaction: the beam radius $r_b$ and  the parameter $\chi_0$. It will be shown below that these two parameters are enough to calculate the relative change of the disruption parameter caused by beamstrahlung. 
In the classical and QED regime Eq.~\eqref{eq.beta0} can be rewritten as follows
\begin{equation}
    \varkappa \approx \alpha \sqrt{2 r_b a_S}\times
    \begin{cases}
        0.67 \chi_0^{3/2}, & \chi_0 \ll 1, \\
        0.37 \chi_0^{1/6}, & \chi_0 \gg 1.
    \end{cases}
\end{equation}

\section{Beamstrahlung dominated regime }

\subsection{Constant force approximation}
Obtaining the general solution of Eqs.~\eqref{eq.dydt}~--~\eqref{eq.dgdt} seems infeasible, thus to make some analytical estimates first we resort to constant force approximation which corresponds to substituting the electron coordinate $\rho$ in the RHS of Eq.~\eqref{eq.dydt} with its initial value $\rho_0 = 1 $. In that case Eqs.~\eqref{eq.dydt}~--~\eqref{eq.dgdt} take the form   
\begin{eqnarray}
    \frac{d^2\rho}{d\tau^2} = -\frac{\mathcal{E} \left( \rho_0 \right)}{\gamma} \eta(\tau) , \\
    \label{eq.app_g}
    \frac{d\gamma}{d\tau} = -P \left( \chi \right),
   \\
   \chi = \chi_0 \frac{\gamma}{\gamma_0} \eta(\tau) .
\end{eqnarray}
According to Eqs.~\eqref{W2}~--~\eqref{W3} in both classical ($\chi \ll 1$) and QED ($\chi \gg 1$) limits, the function $P$ can be approximated as a power function of $\chi$
\begin{eqnarray}
    \label{eq.I}
    P(\chi) = 
    \begin{cases}
        P_C(\chi) \approx 0.67 \alpha a_S \chi^2, & \chi \ll 1, \\
        P_Q(\chi) \approx 0.37 \alpha a_S \chi^{2/3}, & \chi \gg 1.
      \end{cases} 
\end{eqnarray}
In that case we can obtain the solution in quadratures
\begin{eqnarray}
    \label{eq.sol1}
    \gamma &=& \gamma_0 \left( 1 - \frac{P_0 (1 - \nu)}{\gamma_0} \int \limits_{0}^{\tau}\eta^\nu(\tau')d\tau' \right)^{\frac{1}{1-\nu}}  \\
    \rho(\tau) &=& \rho_0 + \dot{\rho}_0 \tau   - \mathcal{E} \left( \rho_0 \right)  \int\limits_{0}^{\tau} d\tau' \int\limits_{0}^{\tau'} \frac{\eta( \tau'' ) }{\gamma(\tau'')} d\tau'',
\end{eqnarray}
where $\nu = 2$ for the classical regime and $\nu = 2/3$ for the QED regime, $P_0 = P \left( \chi_0 \right)$, $\dot{\rho}_0 = \dot{\rho} (\tau = 0 )$.

Let us analyze the obtained solution for a uniform beam $\eta_x =\eta_r =\eta = 1$ for which $\mathcal{E}(\rho) = \rho$. In this case all the integrals can be calculated explicitly. In particular, we get the following solutions for $\gamma$ and $\rho$

\begin{eqnarray}
\label{eq.gamma_const}
\gamma (\tau) &= &\gamma_0\times
    \begin{dcases}
        \left( 1 + \varkappa \frac{\tau}{\tau_{D_0}} \right)^{-1}, & \chi \ll 1, \\
        \left( 1 -  \frac{\varkappa }{3}\frac{\tau}{\tau_{D_0}}  \right)^{3}, &  \chi \gg 1,
    \end{dcases} \\
\label{eq.rho_const}
\rho (\tau) &=&  1 - \frac{\tau^2 }{\tau_{D_0}^2} \times
    \begin{dcases}
        1 + \frac{\varkappa }{3}\frac{\tau}{\tau_{D_0}} , & \chi \ll 1, \\
        \left( 1 -  \frac{\varkappa }{3}\frac{\tau}{\tau_{D_0}}  \right)^{-1}, &  \chi \gg 1,
    \end{dcases}
\end{eqnarray}
where 
$\dot\rho_0 = 0$ is assumed.

It is interesting to note that the dependence of the electron energy on time is identical in terms of $\tau_{D_0}$ for both classical and QED regimes at the beginning of the interaction ($0<\tau \ll \tau_{D_0}$)
\begin{equation}
\label{time_loss1}
\gamma (\tau) \approx \gamma_0 \left( 1 -  \varkappa \frac{\tau } {\tau_{D_0}}  \right).
\end{equation}  
If we introduce the time $\tau_{\gamma}$ after which the electron energy halves because of beamstrahlung, then this time is about $1.6$ times smaller in the classical regime than in the QED regime,
\begin{eqnarray}
\label{time_loss2}
\tau_{\gamma} (\chi \ll 1) &=&  \tau_{BS} ,
\\ 
\tau_{\gamma} (\chi \gg 1) &= &   3 \left( 1 - 2^{-1/3} \right) \tau_{BS}.
\end{eqnarray}  
This is the expected result as the beamstrahlung losses according to the classical expression are greater than that according to the quantum one.


The beamstrahlung-affected disruption time can be found from the condition $\rho(\tau = \tau_D) = 0 $. When we use the relation $D\propto \tau_D^{-2}$ between disruption parameter and disruption time, and when we further consider the case that beamstrahlung sets the timescale of the interaction, $\varkappa\gg1$, then the expression for the disruption parameter including beamstrahlung takes the form
\begin{equation}
    D \approx D_0 
    \begin{cases}
    \left( \varkappa /3 \right)^{2/3} , & \chi \ll 1, \\
    \left( \varkappa /3 \right)^2, &  \chi \gg 1.
    \end{cases}
    \label{d1}
\end{equation}
In virtue of Eq.~\eqref{eq.beta0} we can rewrite Eq.~\eqref{d1}  in terms of $r_b$ and $\chi_0$ as follows
\begin{equation}
    D \approx 
    D_0  \begin{cases}
        2.4 \sqrt[3]{r_b[\upmu \text{m}]}\ \chi_0, & \chi \ll 1, \\
        4.2 \ r_b[\upmu \text{m}]\ \chi_0^{1/3}, & \chi \gg 1. \\
    \end{cases}
\end{equation}


\begin{figure}
    \includegraphics[width=85mm]{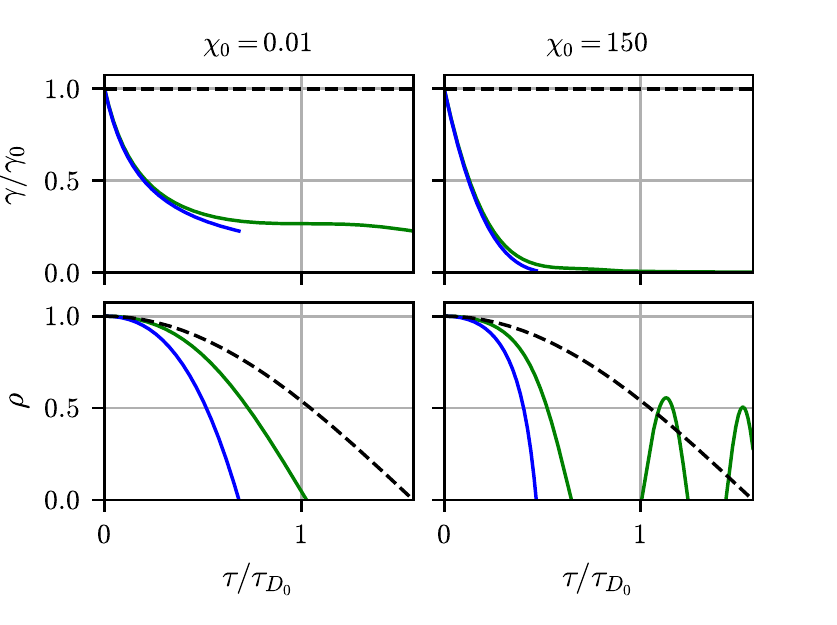}
    \caption{\label{fig.motionless} Comparison of the approximate solution~\eqref{eq.gamma_const}~--~\eqref{eq.rho_const} (blue line) with the numeric solution of Eqs.~\eqref{eq.dydt}~--~\eqref{eq.dgdt} (green line) for $\varkappa_0=5$. $\chi_0=0.01$ for the left column and $\chi_0=150$ for the right column. Black dashed line represents solution of Eq.~\eqref{eq.dydt} with constant value of $\gamma$.}
\end{figure}

Figure~\ref{fig.motionless} shows that while both solutions in quantum and classical regimes describe energy loss quite well, the particle trajectory according to this solution diverges from the real trajectory quite strongly and overestimates beam disruption, thus this simple model can serve only for rough estimates of the disruption parameter, which can be sufficient in cases when only its order of magnitude is of interest.

\subsection{Corrections to the model}\
\label{cor}

The accuracy of the analytical model can be greatly improved by two modifications.
First, we use the mean transverse coordinate in the RHS of Eq.~\eqref{eq.dydt} instead of its initial value $\rho_0$,
 \begin{eqnarray}
\frac{d^2\rho}{d\tau^2} = -\frac{\mu}{\gamma} , \\
\label{rho2}
\frac{d\gamma}{d\tau} = -P \left( \mu \chi_0 \frac{\gamma}{\gamma_0} \right), \\
\mu \equiv \frac{1}{\tau_D} \int^{\tau_D}_{0} \rho \left( \tau' \right) d \tau' < 1. 
\end{eqnarray}
And second, we stitch the solutions in the QED and classical regimes at some time instance $\tau_1$ at which the particle $\chi$ parameter reaches some threshold value $\chi_1 \sim 1$, if its initial value was large enough, i.e $\chi_0 > \chi_1$. So for $\tau < \tau_1$ the equations of motion have the following solution 
\begin{eqnarray}
    \label{eq.gamma_Q}
    \gamma_Q(\tau) &=& \gamma_0 \left(1 - \tilde\varkappa_0 \frac{\tau}{\tau_{D_0}} \right)^{3} ,\\
    \label{eq.rho_Q}
    \rho_Q(\tau) &=& 1 - \frac{\tau^2}{\tau_{D_0}^2}\left( 1-\tilde\varkappa_0 \frac{\tau}{\tau_{D_0}} \right)^{-1}, \\
    \tilde\varkappa_0 &=& \sqrt{\frac{2}{9\gamma_0\mu}}P_Q(\mu\chi_0).
\end{eqnarray}
Please note that the variable $\tilde\varkappa_0$ (and also $\tilde\varkappa_1$, see the next equations) includes an additional factor $1/3$ as compared with the definition of $\varkappa$ in Eq.~\eqref{eq.condition}. This re-definition simplifies later calculations. The time instance $\tau_1$ is found from the condition
\begin{eqnarray}
    \label{eq.chi1}
    \chi = \chi_0 \frac{\gamma_Q(\tau_1) \rho_Q(\tau_1)}{\gamma_0 } \equiv \chi_0 \frac{\gamma_1 \rho_1}{\gamma_0 } = \chi_1.
\end{eqnarray}
From this, the ratio $\gamma_1/\gamma_0$ can be found as follows
\begin{equation}
    \frac{\gamma_1}{\gamma_0} = \frac{\chi_1}{\chi_0} \frac{1}{\rho_1} \equiv \frac{\zeta}{\rho_1},
\end{equation}
where we introduced $\zeta=\chi_1/\chi_0$.

For $\tau > \tau_1$, the classical formulas are used so that the solution of the equation of motion reads
\begin{eqnarray}
    \label{gcl}
    \label{eq.gamma_C}
    \gamma_C(\tau) &=& \gamma_1 \left(1+ 3\tilde\varkappa_1\frac{\tau - \tau_1}{\tau_{D_1}} \right)^{-1} ,\\
    \label{eq.rho_C}
    \nonumber \rho_C(\tau) &=& \rho_1 + \dot{\rho}_1 \frac{\tau - \tau_1}{\tau_{D_1}} - \\ 
    &-& \frac{(\tau-\tau_1)^2}{\tau_{D_1}^2}\left(1+\tilde\varkappa_1\frac{\tau - \tau_1 }{\tau_{D_1}}\right), \\
    \tau_{D_1}  &=& \sqrt{2 \gamma_1}, \\
    \tilde\varkappa_1 &=& \sqrt{\frac{2}{9\gamma_1\mu}}P_C(\mu\chi_1) , \\
    \nonumber \dot{\rho}_1 &=& \tau_{D_1} \dot{\rho}_Q (\tau = \tau_1) = \\
    &=& - \sqrt{\frac{\zeta}{\rho_1}}\frac{\tau_1 }{\tau_{D_0}}\frac{2-\tilde\varkappa_0 \frac{\tau_1}{\tau_{D_0}}}{\left( 1 - \tilde\varkappa_0 \frac{\tau_1}{\tau_{D_0}} \right)^2}, 
\end{eqnarray}

Now the disruption time can be calculated from the condition $\rho_C(\tau_D) = 0$ which can be explicitly solved but the expression is too cumbersome to include it here. Instead it can be estimated with a bit less bulky expression (see Appendix~\ref{app.Pinching} for derivation)
\begin{eqnarray}
    \label{eq.est0}
    \frac{\tau_D}{\tau_{D_0}} = \tau_1 + \tau_2 \sqrt{\frac{\zeta}{\rho_1}} ,\\
    \label{eq.est1}
    \begin{split}
        \tau_1 = \text{min}\left \{ \frac{1 - \zeta ^{1/3}}{\tilde\varkappa_0}, \right. \\ \left. \frac{\tilde\varkappa_0 \left( 1 - \zeta \right)}{2}  \left( \sqrt{1 + \frac{4}{\tilde\varkappa_0^2 \left( 1 - \zeta \right)}} - 1 \right) \right \},
    \end{split}  \\
    \label{eq.est2}
    \tau_2 = \text{min}\left\{ \sqrt[3]{\frac{\rho_1}{\tilde\varkappa_1}}, \tau' - \frac{\tilde\varkappa_1 \tau'^3}{\dot\rho_1 + 2\tau' \left( 1 + \tilde\varkappa_1 \tau' \right)} \right\} ,\\
    \nonumber 
    \tau' = \sqrt{\rho_1 + \frac{\dot\rho_1^2}{4}} - \frac{\dot\rho_1}{2}.
\end{eqnarray}
For the case $\chi_0 < \chi_1$, $\tau_1 \equiv 0$ and $\chi_1$ has to be substituted with $\chi_0$ in $\tau_2$.

Analogously to Eq.~\eqref{eq.beta0} the parameter governing significance of beamstrahlung can be expressed as follows
\begin{equation}
    \tilde\varkappa = \alpha \sqrt{ \frac{2}{9} r_b a_S} \times
    \begin{dcases}
        {\left( \mu\chi_0 \right)}^{2/3} ,\; \chi_0 < \chi_1, \\
        {\left( \mu\chi_0 \right)}^{1/6} ,\; \chi_0 > \chi_1
    \end{dcases}
\end{equation}
and it can be shown that $\tilde\varkappa_1$ can be expressed in terms of $\tilde\varkappa$.
The beamstrahlung-affected disruption parameter in the corrected model is
\begin{equation}
    D = {D_0} \left(\frac{\tau_{D_0}}{\tau_D}\right)^2.
\end{equation}
According to Eqs.~\eqref{eq.est1} and~\eqref{eq.est2}, $D$ can be expressed as being explicitly dependent only on two initial parameters: the beam radius $r_b$ and the value of $\chi_0$. This allows us to scan over only a two-dimensional map of parameters to calculate the value of $D$ from results of full 3D QED-PIC simulations.

Note that although $\mu$ should be calculated in a self-consistent way from the solution obtained above, numeric analysis shows that the value of $\mu$ is close to $0.5$. So to actually find an analytical solution, we treat $\mu$ as a free parameter which we set to $0.5$. This is also justified by the fact that varying $\mu$ in the range $0.3 - 0.7$ does not significantly alter the final value of the disruption parameter.

\section{Interaction of long beams }

In this section we discuss the interaction of long uniform beams of oppositely charged particles when the number of the betatron oscillations is large $\sigma_x / (c \tau_{D_0}) \gg 1$. In the case of electron-positron interactions we are again focuse on the particle dynamics at the front of the beams. However in the case of electron-proton beams, we can neglect the transverse motion of protons because of their large mass and one can explore the dynamics of electrons located in any part of the beam. For simplicity we consider the interaction of uniform beams $\eta_x =\eta_r =\eta = 1$, for which $\mathcal{E}(\rho) = \rho$.  

It is convenient to introduce the following variables 
\begin{eqnarray}
a^{2} & = & \rho^{2}+\gamma\left(\frac{d\rho}{d\tau}\right)^2,\label{b}\\
\phi & = & \arctan(\frac{d\rho}{d\tau}\frac{\sqrt{\gamma}}{\rho}) ,
\end{eqnarray}
where $a$ and $\phi$ are the amplitude and the phase of the betatron
oscillations ($\rho = a \cos\phi $).
In the new varibales, Eq.~\eqref{eq.app_g} takes the form 
\begin{eqnarray}
\frac{da}{dt} & = & -\frac{a}{2\gamma}\sin^{2}\phi\;P\left( \frac{\chi_0}{\gamma_0} a \gamma  |{\cos\phi} | \right).\label{rho}
\end{eqnarray}

To calculate slowly the varying component of the betatron amplitude, $A = \left\langle a\right\rangle $,
we average Eq.~\eqref{rho} over $\phi$ and neglect
the contribution of the fast varying component of $a$ and $\gamma$
\begin{eqnarray}
\frac{dA}{d\tau} & = & -\frac{A}{2\bar{\gamma}}f_{1}\left(\frac{\chi_0}{\gamma_0} A \bar{\gamma} \right),\label{dst-2}\\
f_{1}(v) & = & \frac{1}{2\pi}\intop_{0}^{2\pi}\sin^{2}\phi\;P\left(v|{\cos\phi} |\right)d\phi\label{f1}\\
\frac{d\bar{\gamma}}{d\tau} & = & -f_{2}\left(\frac{\chi_0}{\gamma_0} A \bar{\gamma} \right),\\
f_{2}(v) & = & \frac{1}{2\pi}\intop_{0}^{2\pi}P\left(v |{\cos\phi}|\right)d\phi ,
\end{eqnarray}
where  $\bar{\gamma} = \left\langle \gamma \right\rangle $. Introducing $\bar{\chi}=\left\langle \chi\right\rangle = \chi_0 A \bar{\gamma} / \gamma_0$
one obtains the system describing the electron dynamics averaged over
the betatron oscillations
\begin{eqnarray}
\frac{d\bar{\chi}}{d\tau} & = & -\frac{\bar{\chi}}{2\bar{\gamma}}\left[f_{1}\left(\bar{\chi}\right)+2f_{2}\left(\bar{\chi} \right)\right],\label{v}\\
\frac{d\bar{\gamma}}{d\tau} & = & -f_{2}\left(\bar{\chi } \right).\label{gf2}
\end{eqnarray}
The system has the constant of motion 
\begin{eqnarray}
\ln \bar{\gamma} - g (\bar{\chi} )  & = & \mathrm{const} \label{integral} , \\
g (v) & = & \int \frac{2f_{2}(v)dv} {v f_{1}\left(v \right)+2 v f_{2}\left(v \right)}.
\end{eqnarray}

In the classical limit ($\chi \ll 1$), one has $f_{2}(v) = 4 f_{1}(v)=P_C(v)/2$
and the constant of motion takes the form
\begin{eqnarray}
    \bar{\gamma}^{-9/8} \bar{\chi}
    & = & \mathrm{const}.\label{integral1}
\end{eqnarray}
It follows from Eqs.~\eqref{gf2} and \eqref{integral1} that
\begin{eqnarray}
    \label{eq.long_gC}
    \bar{\gamma} & = & \gamma_{0} S(\tau)^{-4/5} ,\label{g(t)}\\
    \label{eq.long_rC}
    \bar \rho & = & \rho_{0} S(\tau)^{1/5},\\
    S(\tau) & = & 1+\frac{5}{8}\left( \varkappa\frac{\tau}{\tau_{D_0}} \right).
\end{eqnarray}

\begin{figure}
	\includegraphics[width=85mm]{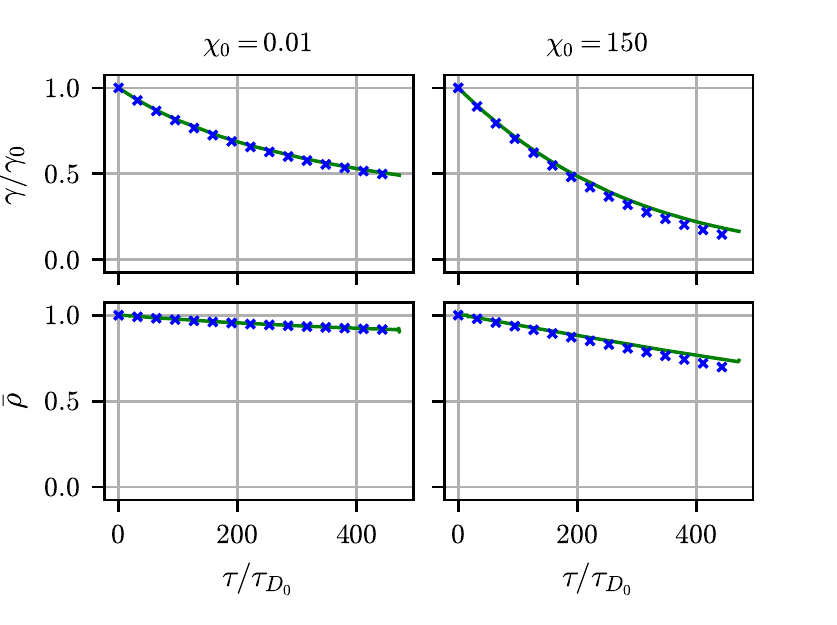}
	\caption{\label{fig.long} Comparison of the approximate solution~\eqref{eq.long_gC}~--~\eqref{eq.long_rC}, ~\eqref{eq.long_gQ}~--~\eqref{eq.long_rQ}(blue crosses) with the numeric solution of Eqs.~\eqref{eq.dydt}~--~\eqref{eq.dgdt} (green line) for $\varkappa_0=0.005$. It is $\chi_0=0.01$ for the left column and $\chi_0=150$ for the right column.}
	\label{long}
\end{figure}

In the QED regime ($\chi\gg1$), one has  
$f_{2}(v)= (8/3) f_1 (v) = \Gamma(5/6) \Gamma^{-1}(4/3) \pi^{-1/2}  P_Q(v)$
and the constant of motion takes the form
\begin{eqnarray}
    \bar{\gamma}^{-19/16}\bar{\chi}
    & = & \mathrm{const}.\label{integral1-1}
\end{eqnarray}
Equations~\eqref{gf2} and \eqref{integral1-1} then yield
\begin{eqnarray}
    \label{eq.long_gQ}
    \bar{\gamma} & = & \gamma_{0} S(\tau)^{24/5},\label{gqed} \\
    \label{eq.long_rQ}
    \bar\rho  & = & \rho_{0} S(\tau)^{9/5},\\
    S(\tau) &=& 1-\frac{5}{24\sqrt\pi}\frac{\Gamma(5/6)}{\Gamma(4/3)}\left( \varkappa\frac{\tau}{\tau_{D_0}} \right) \\
    \nonumber &\approx& 1 - 0.149\left( \varkappa\frac{\tau}{\tau_{D_0}} \right).
\end{eqnarray}
Figure~\ref{fig.long} demonstrates the numerical solution of Eqs.~\eqref{eq.dydt}-\eqref{eq.c2}
and the analytical result given by Eqs.~\eqref{g(t)},~\eqref{gqed} for $\gamma(\tau)$ which are in a good agreement.

\section{PIC simulations}

\begin{figure*}
    \includegraphics[width=165mm]{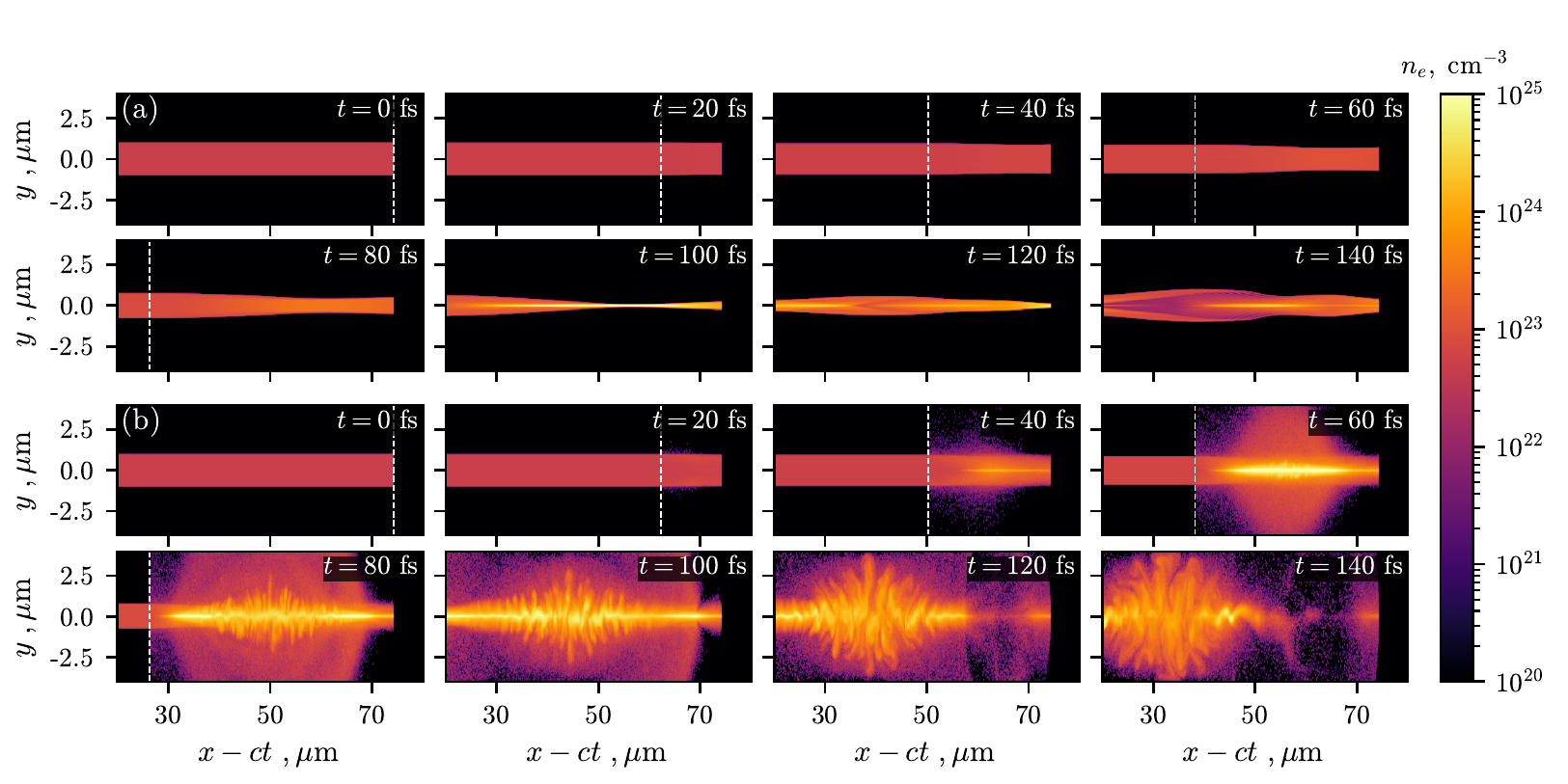}
    \caption{\label{fig.densities} Electron density distribution at different time instances in PIC simulation with parameters $r = 1\ \upmu \text{m}$, $\chi = 10$. White dashed line indicates position of the front of the counter propagating positron beam. Simulation (a) without and (b) with account of QED processes.}
\end{figure*}

\begin{figure*}
    \includegraphics[width=165mm]{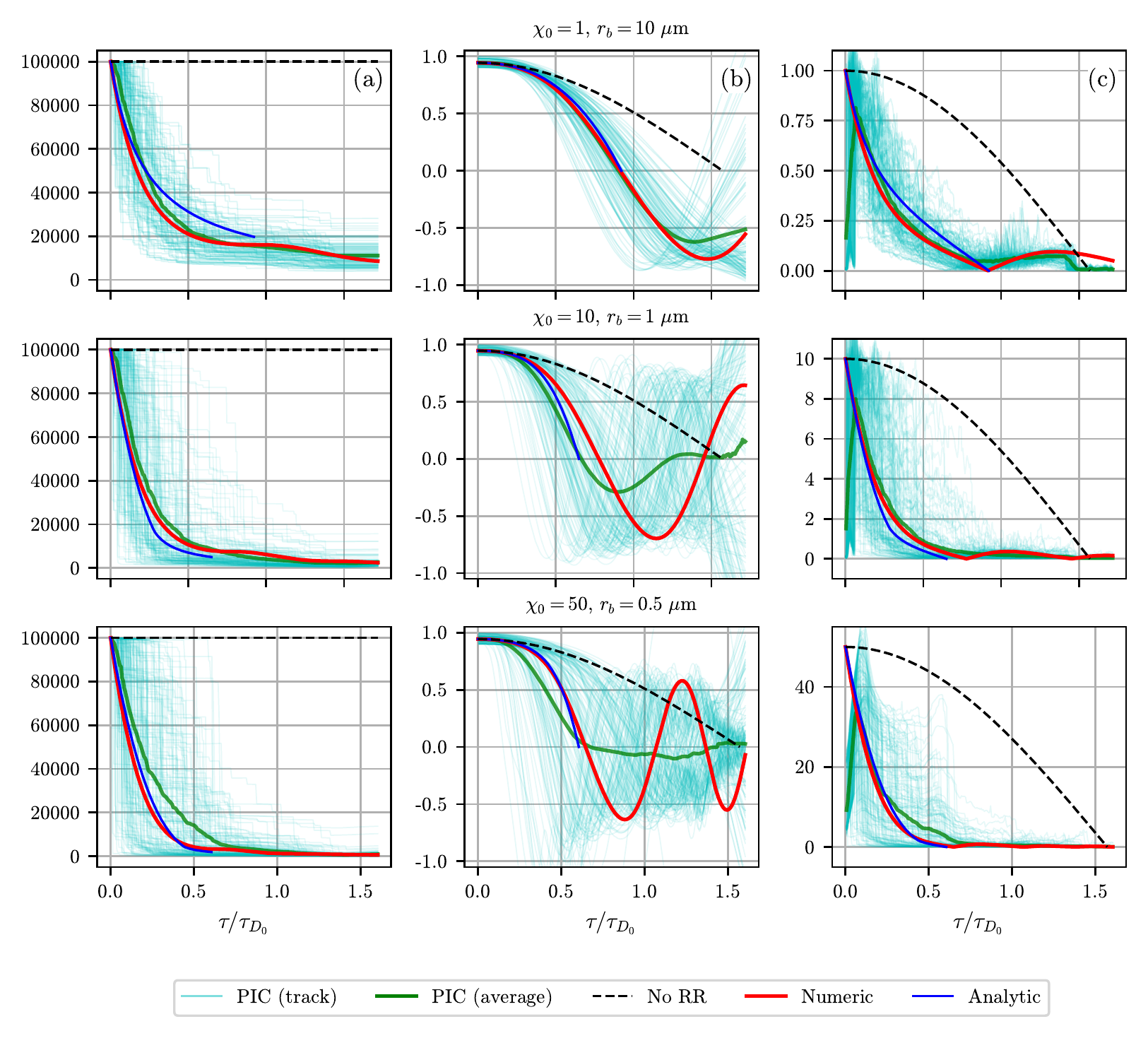}
    \caption{\label{fig.tracks} Electron dynamics in the field of a counter propagating positron beam. (a) Electron energy, (b) displacement from the beam axis and (c) value of $\chi$ parameter as functions of time. Pale cyan lines correspond to individual particles in PIC simulation, green line is an average over particles in PIC simulation, red line represents numerical solution of Eqs.~\eqref{eq.dydt}~--~\eqref{eq.dgdt}, black dashed line represents solution of Eq.~\eqref{eq.dydt} with constant value of $\gamma$, which corresponds to neglecting beamstrahlung, and blue line represents approximate analytical solution~\eqref{eq.gamma_Q}~--~\eqref{eq.rho_Q},~\eqref{eq.gamma_C}~--~\eqref{eq.rho_C}. Different rows correspond to different initial parameters $r_b$ and $\chi_0$.}
\end{figure*}

\begin{figure*}
    \includegraphics[width=165mm]{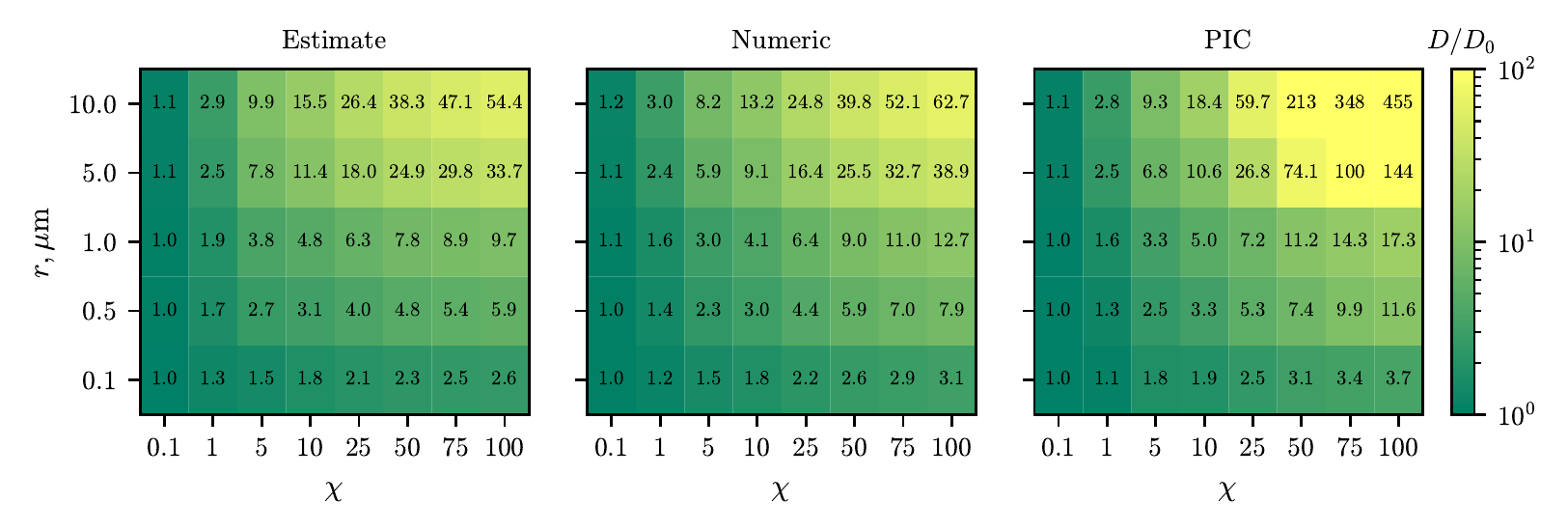}
    \caption{\label{fig.disruption} Value of the ratio $D/ D_0$ (left) calculated according to Eqs~\eqref{eq.est0}~--~\eqref{eq.est1}, (middle) computed from numerical solution of the Eqs.~\eqref{eq.dydt}--\eqref{eq.dgdt}, (right) obtained from results of 3D QED-PIC simulations.}
\end{figure*}

To confirm the prediction of the model developed in Sec.~III. we performed 3D QED-PIC simulations using the QUILL code~\cite{QUILL}, which enables modelling of the QED effects via the Monte-Carlo method. Choosing $x$ as the axis of beam propagation, the simulation parameters were $\Delta t = 0.6 \Delta x$, $\Delta y = 2.5 \Delta x = r_b / 20$. For all performed simulations the resulting time-step $\Delta t$ was much smaller than the average delay between consecutive QED processes, i.e. $W \Delta t \ll 1$ where $W$ is the total probability of some QED process (emission of the gamma-quant or birth of the electron-positron pair). A hybrid FDTD scheme~\cite{samsonov2020superluminal} was used for the numerical solution of  Maxwell's equations and the Vay pusher~\cite{vay2008simulation} was used to push the particles. Simulations were also performed using the VLPL code~\cite{pukhov1999three,NDFX,PhysRevE_94_063204} in combination with the dispersionless RIP solver~\cite{Pukhov2019}. Differences between the results of the simulations using two different codes were insignificant. Figure~\ref{fig.densities} shows an example of such a simulation (see Supplemental material~\cite{Supplemental} for a corresponding video). It shows that at $\chi_0 = 10$ abundant creation of secondary electrons and positrons occurs, which is an evidence of QED cascading. As this process does not effect motion of the beam particles at the front, formation and development of such cascade is not discussed in detail. Also note that development of transverse kink instability is triggered in simulations with QED processes taken into account. This is probably due to the fact that QED processes are stochastic and thus lead to perturbation of the initially symetrical particles distribution acting as a seed for kink instability.

We performed a set of simulations with varying initial radius $r_b$ and $\chi_0$ of the beam particles. The length of the beam was chosen in such a way that for each simulation the uncorrected disruption parameter, i.e. $D_0$, was equal to 10. For each simulation we were tracking several hundreds of the particles located at the front and periphery of the electron beam to calculate the mean time of crossing the beam axis. 
Examples of individual tracks, numerical solution of the Eqs.~\eqref{eq.dydt}~--~\eqref{eq.dgdt} and approximate analytical solution are shown in Fig.~\ref{fig.tracks}.
For each pair of values $r_b$ and $\chi_0$ we performed a simulation with QED processes (Breit-Wheeler and nonlinear Compton scattering) and a reference simulation in which these processes were turned off artificially. By comparing the mean disruption times in these two simulations we are able to calculate the disruption parameter for a wide range of parameters. A map of the value of $D/ D_0$ obtained from PIC simulations is given in Fig.~\ref{fig.disruption}, together with the estimate from Eqs.~\eqref{eq.est0}~---~\eqref{eq.est2} (in which we used $\mu=0.5$, $\chi_1=1$) and with the result from the numerical solution of Eqs.~\eqref{eq.dydt}--\eqref{eq.dgdt}.

It is important to note that for large values of $r_b$ and $\chi_0$ we introduced a different numeric criteria for  the calculation of disruption. This is due to the fact that in such simulations energy loss due to beamstrahlung is so strong that after some time, beam particles are no longer relativistic and their longitudinal velocity becomes comparable to their transverse velocity so eventually the particles stop their directional motion and start spinning without crossing the beam axis (see Suplemental material~\cite{Supplemental}). In such cases instead of the time of reaching the beam axis, we used  the mean time it takes the particles to decelerate down to $0.5c$. As our analytical model assumes that longitudinal velocity is always larger than the transverse one it cannot be applied in these cases.

We did not perform PIC simulations of the beam-beam interaction in the regime when beamstrahlung takes many betatron oscillations to significantly decrease particles energy ($\varkappa \ll 1$) due to several reasons. First, such simulations would take significantly more time. And second, as this regime is mostly related to the electron-proton interaction during which more massive proton bunch does not deform much, such interaction can be sufficiently simulated by a single electron in a given field of the proton bunch, which was done in Sec.~IV.

\section{Conclusions}

The beam dynamics  during the interaction of beams with opposite charges is studied in the beamstrahlung dominated regime.  It turns out that beam radius $r_b$ and $\chi_0$ are the key parameters determining  the regime of beamstrahlung.  For a uniform beam, the parameter $\chi_0$ can be calculated from the beam density $n_e$ or total beam charge $Q$ as follows
\begin{eqnarray}
\label{eq.chi_Q}
\chi_0 & \approx  & 5.3\; \frac{\varepsilon_b[100\; \text{GeV}]\ Q[\text{nC}]}{r_b[\upmu \text{m}]\; \sigma_x [\upmu \text{m}]} \\
\label{eq.chi_ne}
& \approx  & 2.67\; \varepsilon_b[100\; \text{GeV}]\; n_e[10^{21}\; \text{cm}^{-3}] \;r_b[\upmu \text{m}]
\end{eqnarray}
where $\varepsilon_b$ is the beam particle energy. 
We demonstrate that the beamstrahlung effect can strongly enhance the beam focusing. The disruption parameter characterizes the degree of beam focusing in the interaction region. According to the constant force approximation model  the ratio of the beamstrahlung-affected disruption parameter to the beamstrahlung-free disruption parameter can be also expressed in terms of the beam parameters. In the classical regime ($\chi_0 \ll 1$),  it reads
\begin{eqnarray}
    D &\approx&  4\times 10^{-3} \frac{Q[\text{nC}]^2}{r_b[\upmu \text{m}]^{8/3}} \\
    &\approx& 8\times 10^{-4} n_e[10^{21} \, \text{cm}^{-3}]^2\sigma_x[\upmu \text{m}]^2 r_b[\upmu \text{m}]^{2/3},
\end{eqnarray}
\begin{eqnarray}
    \frac{D}{D_0} &\approx&  22.1\; \frac{\varepsilon_b [100\; \text{GeV}]\; Q[\text{nC}]}{r_b[\upmu \text{m}]^{2/3} \;\sigma_x [\upmu \text{m}]} \\
    &\approx& 8.9\; \varepsilon_b [100 \; \text{GeV}] \;n_e[10^{21} \;\text{cm}^{-3}] \;r_b[\upmu \text{m}]^{2/3},
\end{eqnarray}
and  in the QED regime ($\chi_0 \gg 1$)
\begin{eqnarray}
    D &\approx&
    7.2\times 10^{-3} \left( \frac{Q[\text{nC}]^2 \sigma_x[\upmu \text{m}]}{\varepsilon_b[100\, \text{GeV}] r_b[\upmu \text{m}]^2} \right)^{2/3}
    \\
    &\approx& 1.4\times 10^{-3}\left( \frac{n_e[10^{21} \, \text{cm}^{-3}]^{2} \sigma_x[\upmu \text{m}]^3 }{\varepsilon_b [100\, \text{GeV}] r_b[\upmu \text{m}]^{-2}} \right)^{2/3} ,
\end{eqnarray}
\begin{eqnarray}
    \frac{D}{D_0} \approx 38.8\left(\frac{r_b[\upmu \text{m}]^{2} \;\varepsilon_b[100 \;\text{GeV}] \;Q[\text{nC}]}{\sigma_x [\upmu \text{m}]}\right)^{1/3} \\
    \approx 15.6\left( \varepsilon_b[100 \;\text{GeV}] \;n_e[10^{21}\; \text{cm}^{-3}] \;r_b[\upmu \text{m}]^4\right)^{1/3} .
\end{eqnarray}
In above expressions we used the following expression for the beamstrahlung-free disruption parameter $D_0$
\begin{eqnarray}
    D_0 &\approx& 1.8\times 10^{-4}\, \frac{Q[\text{nC}]\;\sigma_x[\upmu \text{m}]}{\varepsilon_b[100\; \text{GeV}] \;r_b[\upmu \text{m}]^2}, \\
    &\approx& 0.9\times 10^{-4} \,\frac{n_e[10^{21}\; \text{cm}^{-3}] \;\sigma_x[\upmu \text{m}]^2}{\varepsilon_b[100 \;\text{GeV}]}.
\end{eqnarray}
 A more accurate value of $D$ can be calculated from the corrected model presented in Sec.~\ref{cor}.  
The developed model can be also extended to more realistic beam profiles  in a straightforward way.  The numerical solution of the  equations of motion in the field of Gaussian and parabolic beams  shows that generally our analytical estimate can still be used  when using the calculation based on  the average value of $\chi$.

The analytical model describing  the interaction of long oppositely charged  beams is also presented. It is assumed that the beam particles perform a number of betatron oscillations along the beam axis. The dependence of the particle energy and its amplitude of betatron oscillations are calculated in  the classical regime as well as in  the QED regime of beamstrahlung. The model prediction is in good agreement with results of numerical simulations.

In order to study disruption effect the dynamics of the electrons located only at the beam front were examined, although the beam-beam interaction is determined by all the particles of the beam. This is especially  the case for intense regime of interaction accompanied be QED cascading. 
Abundant production of the secondary particles in such a cascade observed in numerical simulations leads to disturbance of the initially symetrical distribution, which serves as a seed for development of the kink instability. This effect requires further study as it may be as limiting for collidier operation as the disruption effect itself. Athough to accurately describe dynamics of the beams as a whole one would need to self-consistently calculate electromagnetic field distribution, which makes the problem much more complicated for analysis.

\begin{figure*}
    \includegraphics[width=165mm]{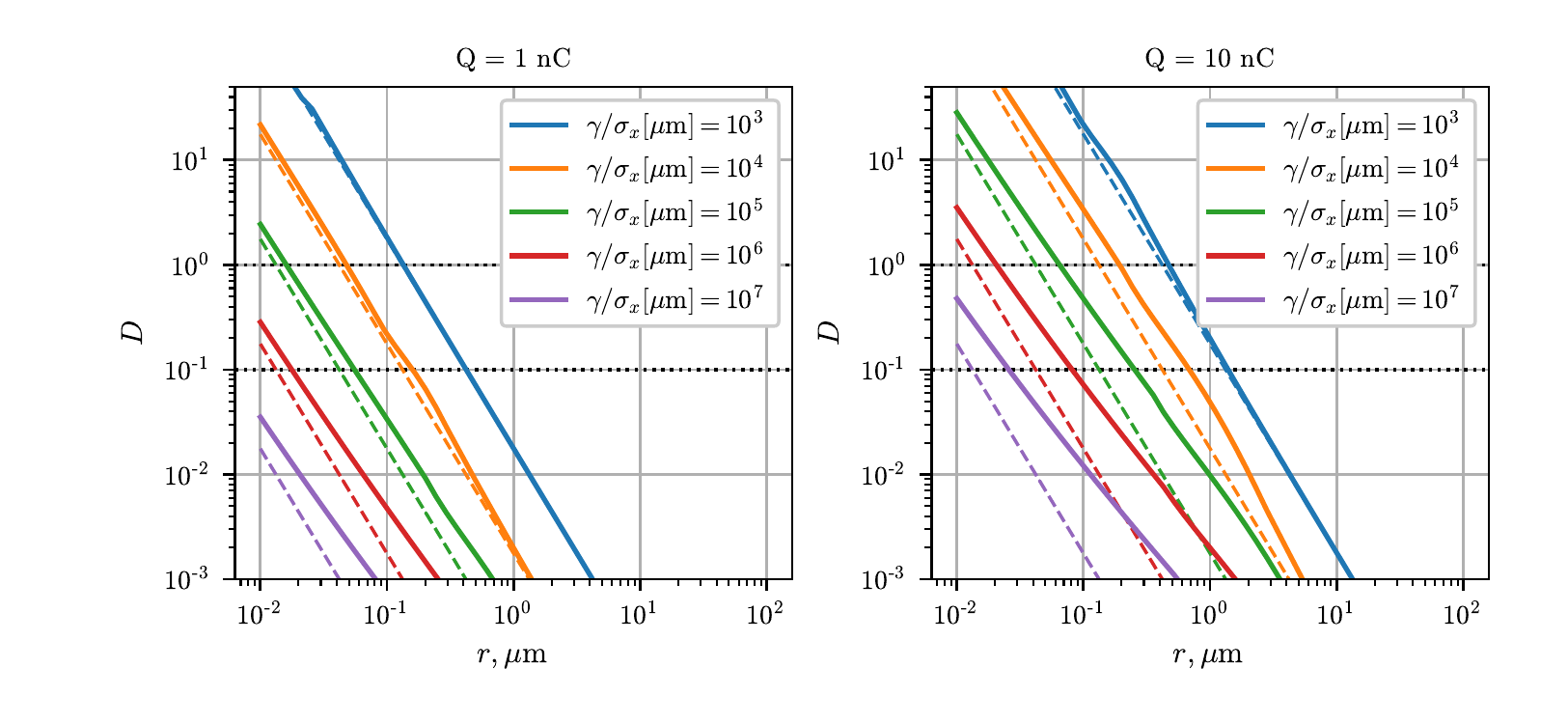}
    \caption{\label{fig.Q_disruption} Disruption parameter calculated with (solid lines) and without (dashed lines) account of beamstrahlung for different beam parameters.}
\end{figure*}

Upcoming colliders such as FACET-II, CLIC, ILC are designed to have controlled beam-beam interaction, i.e. $D \ll 1$ and even accounting beamstrahlung does not raises disruption parameter above unity (see Fig.~\ref{fig.Q_disruption}). Fig.~\ref{fig.Q_disruption} also shows that collision of beams with quite a large total charge ($> 10$ nC) with small radii can be significantly altered by beamstrahlung. Another interesting trend that can be observed is that although increasing particle energy and/or decreasing beam length (while preserving the same total charge) decreases disruption parameter, at the same time it increases significance of beamstrahlung. Overall ratio $D/ D_0$ can be used to determine whether radiation effects are significant or not. 
Several experimental setups were proposed aiming to investigate nature of the radiation reaction and test models commonly used to describe it~\cite{yakimenko2019prospect, tamburini2020efficient}. In such setups it is crucial to maximize  the effect of radiation friction. This can be done by exploiting round beams instead of flat ones (focused much stronger along one direction than along another one). 

\begin{acknowledgments} 
	
This work was supported by the Russian Foundation for Basic Research (project No. 20-52-12046), Foundation for the advancement of theoretical physics and mathematics 'BASIS' (Grant No.~19-1-5-10-1) and by the Deutsche Forschungsgemeinschaft (DFG) under project number 430078384. The authors gratefully acknowledge the Gauss Centre for Supercomputing e.V. (www.gauss-centre.eu) for funding this project (qed20) by providing computing time on the GCS Supercomputer JUWELS at J\"ulich Supercomputing Centre (JSC).
	
\end{acknowledgments}

\appendix

\section{Estimation of pinching time}
\label{app.Pinching}
\subsection{Estimation of $\tau_1$}
To find an estimate for the time instance $\tau_1$ defined in Eq.~\eqref{eq.chi1}, let us consider the following equation on $x$
\begin{equation}
    \label{app.zeta}
    k_1 = \left( 1 - \frac{x^2}{1 - k_2 x} \right) \left( 1 - k_2 x \right)^3 .
\end{equation}
As both factors decrease with $x$ it is evident that $x < x_{1,2}$, where
\begin{align}
    k_1 = {\left(1 - k_2 x_1\right)}^3 ,\\
    k_1 = 1 - \frac{x_2^2}{1 - k_2 x_2}.
\end{align}
These equations have the following solutions
\begin{align}
    x_1 = \frac{1-\sqrt[3]{k_1}}{k_2} \\
    x_2 = \frac{k_2(1-k_1)}{2} \left( \sqrt{1 + \frac{4}{k_2^2 (1-k_1)}} - 1 \right).
\end{align}
 Finally, an approximate solution of Eq.~\eqref{app.zeta} can be found as $x = \text{min}\{x_1, x_2\}$. To find $\tau_1$, we perform the following substitution
\begin{eqnarray}
    x \rightarrow \frac{\tau_1}{\tau_{D_0}} ,\ 
    k_1 \rightarrow \frac{\chi_1}{\chi_0} = \zeta ,\ 
    k_2 \rightarrow \tilde{\varkappa}_0.
\end{eqnarray}

\subsection{Estimation of  $\tau_2$}
To estimate the disruption time from the condition $\rho_C(\tau_D)=0$ [$\rho_C$ is defined in Eq.~\eqref{eq.rho_C}], let us consider the following equation on $x$
\begin{equation}
    \label{app.rho}
    0 = k_1 + k_2 x - x^2 \left( 1 + k_3 x \right).
\end{equation}
For large values $k_3$, a rough estimate for solving this equation is
\begin{align}
    x_1 = \sqrt[3]{\frac{k_1}{k_3}}.
\end{align}
For smaller values $k_3$, we can first find a solution by setting $k_3=0$, i.e.
\begin{align}
    0 = k_1 + k_2 x' - x'^2.
\end{align}
The above equation has the solution
\begin{equation}
    x' = \frac{k_2}{2} + \sqrt{k_1 + \frac{k_2^2}{4}} .
\end{equation}
By assuming that the solution of Eq.~\eqref{app.rho} is only slightly different from $x'$, i.e. $x=x'+x''$, we can expand this equation in $x''$ and keep only linear terms:
\begin{align}
    k_2 x'' + 2 x'' x' (1 + k_3 x') - k_3 x'^3 = 0.
\end{align}
From this we obtain that $x$ can be approximated in the following way
\begin{align}
    x_2 = x' - \frac{k_3 x'^3}{k_2 + 2 x' (1 + k_3 x')}
\end{align}
And finally we choose the smallest of $x_{1,2}$, i.e. $x~=~\text{min}\{x_1, x_2\}$.
To find $\tau_2$, we perform the following substitution
\begin{eqnarray}
    x \rightarrow \frac{\tau_2}{\tau_{D_1}} ,\ 
    k_1 \rightarrow \rho_1 ,\ 
    k_2 \rightarrow \dot{\rho}_1 ,\ 
    k_3 \rightarrow \tilde{\varkappa}_1. 
\end{eqnarray}
\bibliography{main}

\end{document}